\documentclass{amsart}
\usepackage{epsfig}
\usepackage{latexsym}
\usepackage{amsmath}
\usepackage{amscd}

\newtheorem{theorem}{Theorem}[section]
\newtheorem{proposition}[theorem]{Proposition}
\newtheorem{lemma}[theorem]{Lemma}
\newtheorem{corollary}[theorem]{Corollary}

\def\R {{\mathcal R}}
\def\Q {{\mathcal Q}}

\def\P {{\mathcal P}}

\newcommand{\PP}{{\mathbb P}}

\title[Random autocatalytic networks]{Random biochemical networks: the probability of self-sustaining 
autocatalysis}
\author{Elchanan Mossel, Mike Steel} 

\address{Elchanan Mossel: Statistics Department, UC Berkeley, CA, USA;
Mike Steel:  Allan Wilson Centre for Molecular Ecology and Evolution, 
University of Canterbury, Christchurch, New Zealand}

\email{mossel@stat.berkeley.edu,  m.steel@math.canterbury.ac.nz}

\subjclass{} 

\date{14 May 2004} 

\keywords{combinatorial chemistry, autocatalysis, discrete random structures}

\begin{document}
\begin{abstract} 
We determine conditions under which a random biochemical system is likely to contain 
a subsystem that is both autocatalytic and able to survive on some ambient `food' source. Such systems have
previously been investigated for their relevance to origin-of-life models. In this paper we extend earlier work, by
finding precisely the order of catalysation required for the emergence of such self-sustaining autocatalytic networks.
This answers questions raised in earlier papers, yet also allows for a more general class of models. We also show that 
a recently-described
polynomial time algorithm for determining whether a catalytic reaction system contains an autocatalytic, self-sustaining
subsystem is unlikely to adapt to allow inhibitory catalysation - in this case we show that the associated 
decision problem
is NP-complete. 

\end{abstract}

\maketitle
{\em Address for correspondence:} 

Mike Steel 

Biomathematics Research Centre,
University of Canterbury

Private Bag 4800,
Christchurch,
New Zealand 

Email: m.steel@math.canterbury.ac.nz 

Fax: 64-3-3642587.

\newpage

\section{Introduction}

The idea that the study of discrete random networks could provide some
insight into the problem of how primitive life might have emerged from
an ambient `soup' of molecules goes back the mid-1980s. This was largely
motivated by the earlier investigation of random graphs, pioneered by
Alfred R{\'e}nyi and Paul Erd{\"o}s in the 1950's and 1960s, which had
revealed the widespread occurrence of `threshold phenomena' (sometimes
also called `phase transitions') in properties of these graphs
(Erd{\"o}s and R{\'e}nyi, 1960).   In the simplest random graph model one has set of vertices (points) and edges are added independently and randomly between pairs of vertices. 
As the probability that any two nodes are jointed by an edge passes certain well-studied thresholds, there is 
typically a fundamental change in various qualitative properties of a
large random graph, such as its connectivity, or the size of the largest
component (see eg. Bollobas, 2001). Extending this approach, Bollobas
and Rasmussen (1989) investigated when a directed cycle would first
emerge in a random directed graph, and how many vertices such a cycle
would contains. They were motivated by the idea that the emergence of a
primitive metabolic cycle was an essential step in the early history of
life, writing ``we want to know when the first catalytic feedbacks
appear, and how many different RNA molecules they involve."  Cohen
(1988) also foresaw the relevance of random graph techniques for modelling primitive
biological processes.

The importance of cycles in early life had also been studied - from a slightly different perspective - by Eigen (1971) and Eigen and Schuster (1979). They proposed a metabolic `hypercycle' as a way of circumventing the so-called `error catastrophe' in the formation of longer strings of nucleotides, first demonstrated by Maynard-Smith (1983). The study of such processes and how they might further evolve into early life has been extensively investigated, using both stochastic and dynamical approaches (eg. Scheuring, 2000; Wills and Henderson, 1997; 
Zintzaras, Santos and Szathm{\'a}ry, 2002). 

The idea that threshold phenomena might help explain some of the mystery surrounding the emergence of life-like systems from a soup of inanimate molecules was developed further by Dyson (1982, 1985) and Stuart Kauffman (1986, 1993). Kauffman considered simple autocatalytic protein networks where amino acid sequences catalyse the joining 
(or `ligation') of shorter sequences, and the cutting (or `cleavage') of
longer sequences.  He calculated that under a simple model of random
catalysation, once the collection of sequences became sufficiently extensive there would inevitably emerge a subsystem of reactions that was both autocatalytic and able to be sustained from an ambient supply of short sequences (such as single or pairs of amino acids). Kauffman realised that simple random graphs and digraphs by themselves do not capture the intricacy of chemical reactions and catalysis. A more complex discrete structure - which has become known as a {\em catalytic reaction system} is required in order to formalize and study the concept of a system of molecules that catalyses all the reactions required for their generation, and which can be sustained from some ambient `food' source of molecules, $F$.  A different  discrete model for self-reproducing systems based on Petri nets has also been developed by Sharov (1991) for investigating the dynamical properties of these systems, but we do not deal with this model here.

Several investigators have developed the studey of catalytic reaction systems  and random autocatalysis (Hordijk and Fontanari, 2003; Hordijk and Steel, 2004; Lohn {\em et al.}, 1998; Wills and Henderson, 1997) though it also has its critics (eg. Lifson, 1997; Orgel, 1992; Maynard Smith and Szathm{\'a}ry 1995) and these criticisms are mainly of two types. Firstly Kauffman invoked overly simplistic and strong assumptions in his analysis - for example he considered just binary sequences (i.e. two amino acids) and assumed that each molecule had the same  fixed probability of catalysing any given reaction. In this paper we make much weaker, and thereby hopefully more robust assumptions in our probabilistic analysis. A second concern is more general - the concept of a `protein-first' start to life is problematic, since proteins, unlike RNA are not able to replicate (for a discussion of this point, part of the so-called `chicken and egg' problem see Lifson, 1997; Maynard Smith and Szathm{\'a}ry 1995; or Penny 2004).  Thus it is quite likely that other sequences besides proteins (such as RNA) may have been part of the first prebiotic systems, and there has been considerable interest from biochemists in the feasibility of an `RNA world' in the early stages of the  formation of life (for a recent survey, see Penny 2004).

At this point there are at least two ways to formalize the concept of a
self-sustaining and autocatalytic set of molecules - the two we study
here are referred to as the RAF (reflectively autocatalytic, and $F$--generated) and
CAF (constructively autocatalytic and $F$--generated) sets. The former was investigated in Steel (2000) and 
Hordijk and Steel (2004). A CAF, which we formalize in this paper is a
slightly stronger notion - it requires that any molecule $m$ that is
involved in any catalysation must already have been built up from
catalysed reactions (starting from $F$).  This concept is perhaps overly
restrictive, since it might be expected that $m$ would still be present
in a random biochemical system in low concentrations initially before
reactions that generate a steady supply of $m$ become established.

For the sequence-based models of the type studied by Kauffman, we determine the degree of catalysation required for a RAF or a CAF to arise.  In Kauffman's model  reactions consist of the concatenation and cutting of sequences up to some maximal (large) length, starting from small sequences of length at most $t$, and each molecule has a certain probability of (independently) catalysing any given reaction. Let $\mu(x)$ denote the average number of (concatenation) reactions that sequence $x$ catalyses, which may depend on $|x|$ the length of $x$.  Then, roughly speaking, our results show that if $\mu(x)/|x|$ is small the probability that the system contains a RAF is small; conversely if $\mu(x)/|x|$ is large the probability the system contains a RAF is close to 1, and indeed in this case there is likely to be a RAF for which all the molecules in the system are involved.  This confirms two conjectures that were posed in Steel (2004) and confirms some trends that were suggested by simulations in Hordijk and Steel (2004). 

Our results for RAFs contrast sharply with the degree of catalysation required for a CAF. In that case each molecule needs to catalyse, on average, some fixed proportion of \underline{all} reactions for a CAF to be likely. That is, the corresponding value of $\mu(x)$ required for a likely occurrence of a CAF is exponentially larger (with $n$) than for a RAF.

We begin this paper by formalizing the concepts of RAF and CAF, and we do so in a more general 
setting than Hordijk and Steel (2004) as we consider the effect of
general catalysation regimes - for example by allowing certain molecules
to inhibit certain reactions. In this case determining whether an
arbitrary catalytic reaction system contains a RAF seems to be
computationally intractable. Indeed  we show that
the decision problem is NP-complete. This contrasts with  the situation
where one allows only positive catalysation; in that case a
polynomial-time algorithm (in the size of the system) for finding a RAF
if one exists was described in Hordijk and Steel (2004). Sections
\ref{secseq} and \ref{seccaf} present the main results concerning the
required growth of $\mu(x)$ with $|x|$ required for RAF and CAR
generation, and in Section~\ref{seccon} we make some concluding
comments, and raise some questions for further investigation. 

Although the assumptions in Kauffman's original paper were quite strong
- for example that each molecule had the same probability of catalysing
any given reaction - in this paper we have been able to weaken some of
these assumptions. The analysis in this paper still ignores
inhibitory catalysis, side reactions that may deplete certain reactants
(Szathm{\'a}ry 2000),  and dynamical aspects of the process (Szathm{\'a}ry and
Maynard Smith 1995) however we hope to extend this analysis in future
work.

\section{Preliminaries and definitions}
\label{secdef}

We mostly follow the notation of Steel (2000) and Hordijk and Steel (2004).  Let $X$ denote a set of 
molecules and $\R$ a set of reactions, where we regard a reaction as an
ordered pairs $(A,B)$ where $A,B$ are subsets of $X$ called the {\em reactants} and {\em products} respectively.
Let $F$ be a distinguished subset of $X$, which can be regarded as some plentiful supply (`food') of reactants.

For $r \in \R$ let  $\rho(r) = A$ and $\pi(r)=B$ and for 
 a set $\R' \subseteq \R$ let $$\rho(\R'):= \cup_{r \in \R'}\rho(r),$$
 $$\pi(\R'):= \cup_{r \in \R'}\pi(r),$$ and 
$${\rm supp}(\R') := \rho(\R') \cup \pi(\R').$$ 

Thus ${\rm supp}(\R')$ denotes the molecules in $X$ 
that are used or produced by at least one reaction in $\R'$.

Given a subset $\R'$ of $\R$ and a subset $X'$ of $X$ the {\em closure} of $X'$ relative to $\R'$, denoted
$cl_{\R'}(X')$ is the (unique) minimal subset $W$ of $X$ that contains $X'$ and that satisfies the following condition for each reaction $(A,B) \in \R'$: 
$$A \subseteq X' \cup W \Rightarrow B \subseteq W.$$
It is easily seen that $cl_{\R'}(X')$ is precisely the set of molecules that can be generated starting from $X'$ and repeatedly applying reactions selected (only) from $\R'$.

Let $\gamma: 2^X \times \R \rightarrow \{0,1\}$ be a {\em catalysation function}.  
The function $\gamma$ tells us whether or not each reaction $r$ can proceed in its environment (eg. be `catalysed')  depending on what other molecules are present.  Thus let $\gamma(A,r) = 1$ precisely when $r$ would be catalyzed if the other molecules in the system comprise the set $A$.  
For example, consider a simple scenario where each reaction $r \in \R$ is catalysed provided that at least one molecule in some set (specific to $r$) is present. We can represent the associated function $\gamma$ as follows -- 
we have a set $C \subseteq X \times R$ (as in Steel, 2000; Hordijk and Steel, 2004) where $(x,r)$ indicates that molecule $x$ catalyses reaction $r$. The catalysation function $\gamma=\gamma_C$ for this simple setting is then defined by
\begin{align*}
\gamma_C(A,r) =\begin{cases} 1, & \mbox{if $\exists x \in A: (x,r) \in C$;}
\\ 0, &
\mbox{otherwise. } \end{cases} \end{align*}

More generally, suppose we have two arbitrary sets $C(+) \subseteq X \times R$  and $C(-) \subseteq X \times R$,
which can represent, respectively the molecules that catalyse and inhibit the various reactions. 
Then a candidate for $\gamma$ is the function $\gamma = \gamma_{C(+),C(-)}$ defined by:
\begin{align} \label{eq:+-}
\gamma_{C(+), C(-)}(A,r) =\begin{cases} 1, & \mbox{if $\exists x \in A: (x,r) \in C(+)$ and there is no $x' \in A: (x',r) \in C(-)$;}
\\ 0, &
\mbox{otherwise. } \end{cases} \end{align} 
Thus $\gamma_{C(+), C(-)}$ allows both catalysation and inhibition.
We find it useful to write 
$A \rightarrow B$ to denote the reaction $(A,B)$. Similarly, we will 
write $A \xrightarrow{C(+),C(-)} B$ 
to denote the reaction $(A,B)$ together with a 
catalysation function that satisfies (\ref{eq:+-}). When the sets $A,B,C$ are 
singletons we will often omit the $\{\}$ symbols.

In case $\gamma$ is monotone in the first co-ordinate (i.e. $A \subset B \Rightarrow \gamma(A,r) \leq \gamma(B,r)$) we
will call $\gamma$ {\em monotone}. Note that $\gamma_C$ is monotone, and that monotone catalytic functions do not allow
inhibition effects.

The triple
$\Q=(X,R,\gamma)$ is called a {\em catalytic reaction system}.

\subsection{Autocatalytic networks}
 Suppose we are 
given a catalytic reaction system $\Q=(X,\R,\gamma)$ and a subset $F$ of $X$.

A {\em reflexive autocatalytic network over F} or $RAF$ for $\Q$ is a non-empty subset $\R'$ of $\R$ for 
which 
\begin{itemize}
\item[(i)] $\rho(\R') \subseteq cl_{\R'}(F)$
\item[(ii)] For each $r \in R', \gamma({\rm supp}(\R'), r) =1$.
\end{itemize}
In addition, to avoid biological triviality, we will also require that any $RAF$ $\R'$ also satisfies the condition
\begin{itemize}
\item[(iii)] $\pi(\R') \not\subseteq F$
\end{itemize}

Thus for $\R'$ to be a RAF, each molecule involved in $\R'$ 
 must be able to be constructed from $F$ by repeated applications of reactions that lie just in $\R'$ (condition (i)) and each reaction in $\R'$ must be catalysed by the system of molecules involved in $\R$ (condition (ii)). This definition is a slight generalization of that given by Hordijk and Steel (2004) to allow for more general catalysation functions $\gamma$ in condition (ii).  Condition (iii) simply ensures that 
any set of reactions that produce only molecules that are already in the food set $F$ does not constitute a RAF.

Next we describe a condition which is somewhat stronger than the RAF requirement.

A {\em constructively autocatalytic network over F} or $CAF$ for $\Q$ is a strictly nested sequence
$\emptyset \neq \R_1 \subset \R_2 \subset \cdots \subset \R_k$, for which 
\begin{itemize}
\item[(i)] $\rho(\R_1) \subseteq F$ and for each $r \in \R_1$, $\gamma(F,r)=1$.
\item[(ii)] For all $i \in \{1,\ldots, k-1\}$, $\rho(\R_{i+1}) \subseteq {\rm supp}(\R_i),$ 
and for each $r \in \R_{i+1}$, $\gamma({\rm supp}(\R_{i}),r) =1$.
\item[(iii)] $\pi(\R_1) \not\subseteq F$.
\end{itemize}

Informally, a CAF is a way to sequentially build up a set of molecules, starting with $F$, and
in such a way that every reaction is catalyzed by at least one molecule that has already
been constructed. Notice that for any catalytic reaction system $\Q$, any set $\R_i$ occurring in a CAF for $\Q$
is also a RAF for $\Q$.

\begin{figure}[ht]
\resizebox{10cm}{!}
{\input{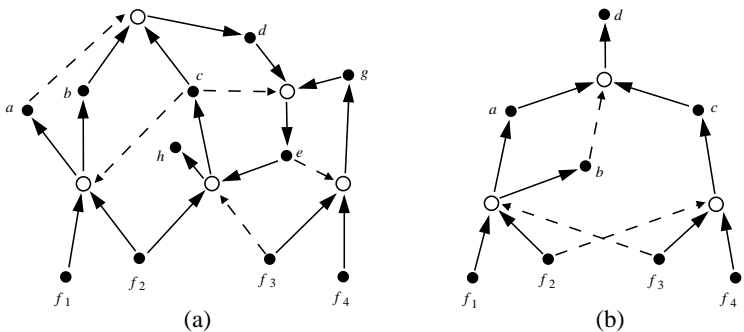}}
\caption{(a) An example of a RAF and (b) a CAF; represented as directed graphs.   Molecules are shown as black nodes, reactions are white nodes, $F= \{f_1, f_2,f_3, f_4\}$ and each (positive) catalysation of a reaction by a molecule is indicated by dashed arc. Solid arcs show the input and output of each reaction.}
\label{example}
\end{figure}

Figure~\ref{example} illustrates these two concepts in the case of simple catalysation (of the form $\gamma=\gamma_C$).


There is a further condition we can impose on a RAF or CAF to make it
more biologically relevant  - namely we may require that a set of
reactions  is capable of constructing complex molecules required for
maintaining certain  biological processes (such as metabolism, error
correction or reproduction). Of course there may be many combinations of complex molecules that suffice to maintain these processes, but we would like $\R'$ to be able to construct at least one of these combinations. 
We can formalize this notion as follows. Suppose $\R'$ is a RAF (respectively 
$R_1 \subset \R_2 \subset \cdots \subset \R_k = \R'$ a CAF) and suppose  
$\Omega \subseteq 2^{X-F}$ is a distinguished collection of subsets of 
molecules.  We say that $\R'$ is an {\em $(\Omega$)--complex RAF}
(respectively an {\em $(\Omega$)--complex CAF}) if
the following condition ($\Omega$C) holds:
\begin{itemize}
\item[($\Omega$C)] If $\Omega \neq \emptyset$ then $C \subseteq \pi(\R')$ for at least one $C \in \Omega$. 
\end{itemize}
We can think of each set $C \in \Omega$ as a suite of complex molecules
that are required for maintaining certain biological processes and
condition ($\Omega$C) requires that the RAF or CAF be capable of constructing at least one such set. 
Note that the definition of an $\Omega$--complex RAF (respectively
$\Omega$--complex CAF) reduces to that of a (simple) RAF or CAF
if we take $\Omega = \emptyset$.

\section{The complexity of determining whether or not $\Q$ has a RAF or a CAF}
\label{seccom}

In Hordijk and Steel (2004) it was shown that, when $\gamma=\gamma_C$,
there is a polynomial-time algorithm to determine if
$\Q$ has a RAF.  However if one allows inhibition also -- by replacing $\gamma_C$ by  $\gamma=\gamma_{C(+), C(-)}$-- 
it is unlikely that any efficient algorithm exists for determining a RAF, by virtue of the following result whose proof is given in the Appendix.

\begin{proposition}
\label{hard}
For arbitrary catalytic reaction systems $\Q= (X,\R,\gamma_{C(+), C(-)})$ and a subset $F$ of $X$ 
the decision problems `does $\Q$ have
a RAF?' is NP-complete. 
\end{proposition}

However for any monotone catalysation function $\gamma$ there is a simple algorithm to determine whether or not
$\Q$ has a CAF,  which is essentially to
let the system `evolve from F'. We describe this now.

\begin{proposition}
\label{algorithm}
Given a catalytic reaction system $\Q= (X,\R,\gamma)$, with $\gamma$ monotone, 
there is a polynomial time algorithm (in $|X|, |\R|$) to
determine whether or not $\Q$ has a CAF.  
\end{proposition}
\begin{proof}
Define a sequence $X_i, \R_i$ for $i \geq 1$ as follows:
$$X_1 = F; \R_1 = \{r \in \R: \rho(r)  \subseteq F, \mbox{ and }  \gamma(F,r)=1\},$$
and for $i \geq 1$ set
$$X_{i+1} = X_i \cup \pi(R_i); 
R_{i+1} = R_i \cup \{r \in \R: \rho(r)  \subseteq X_i \mbox{ and } \gamma(X_i, r) =1 \}.$$
Then provided $\R_1 \neq \emptyset$ the sequence $\R_1 \subseteq \R_2 \cdots \subseteq \R_k$ (for any $k \geq 1$) is
a CAF for $\Q$.  If $\R_1$ is empty, then clearly $\Q$ has no CAF.  
\end{proof}

\section{Random sequence-based models}
\label{secseq}

In this section we take $X=X(n)$, the 
set of sequences of length at most $n$ over the alphabet set $\{0,1,\ldots, \kappa-1\}$.
Let $F$ be a distinguished (small) subset of $X(n)$; in this paper we will take
$F = X(t)$ for a fixed value of $t$ (often a value such as $t=2$ has
been taken in earlier papers).   For a sequence $x \in X(n)$ we will let 
$|x|$ denote its 
length.  
Let $\R(n)$ denote the set of all ordered pairs $r = (A,B)$ where,
for some $a,b,c \in X$ for which $c = ab$ (= the concatenation of $a$ and $b$) either
$A = \{a,b\}$ and $B = \{ab\}$ - in which case we call $r$ a {\em forward reaction} 
or $A = \{ab\}$ and $B= \{a,b\}$ - in which case we call $r$ a {\em backward reaction}.
We may think of the pair $r=(\{a,b\},c)$ as 
representing the ligation reaction 
$$a +b \rightarrow c;$$ and the pair $r=(\{c\},\{a,b\})$ as representing the
cleavage reaction $$c \rightarrow a+b.$$
We will let $\R_{+}(n)$ and $\R_{-}(n)$ denote the (partitioning) subsets of $\R(n)$ consisting of the forward 
and backward reactions, respectively. 
 
 Note that we have 
 \begin{equation} \label{eq:sizeX}
 x_n := |X(n)| = \kappa + \kappa^2 + \cdots + \kappa^n = \frac{\kappa^{n+1} - \kappa}{\kappa-1},
 \end{equation} 
which is  the total number of sequences of length at most $n$,
 and 
 \begin{equation} \label{eq:sizeR}
 r_n := |\R_{+}(n)| =  \left( \kappa^2 + 2\kappa^3+...(n-1)\kappa^n \right) = 
 \frac{(n-1) \kappa^{n+2} -  n \kappa^{n+1} + \kappa^2}{(\kappa-1)^2},
 \end{equation}
which is the total number of forward reactions that construct sequences of length at most $n$.
 We will often below use the fact that, for all $n \geq 1$,  
 \begin{equation} \label{eq:size_ration1}
 1 - O\left(\frac{1}{n}\right) \leq \frac{r_n}{n x_n} \leq 1
 \end{equation}
(the notation $f(n) = g(n) + O(\frac{1}{n})$ means $|f(n)-g(n)| \leq K/n$ for some constant $K$ for all $n \geq 1$).

We study a catalysation function $\gamma$ obtained by setting $\gamma= \gamma_C$ where $C$ is some random assignment of catalysation (i.e. pairs $(x,r)$)
that is subject to the following requirements:
\begin{itemize}
\item[(R1)]
The events $( (x,r) \in C : x \in X(n), r \in \R_{+}(n))$ are independent. 
\item[(R2)]
For each sequence $x \in X(n)$ and reaction $r \in \R_{+}(n)$,
the probability $\PP[(x,r) \in C]$ depends only on $x$.
\end{itemize}

This model is more general than that described in Kauffman (1993),  Steel (2000) or Hordijk and Steel (2004) for several reasons -- 
it allows different catalysation probabilities for forward and backward reactions,
it allows dependencies involving the catalysation of 
backward reactions, and the catalysation ability of a molecule can vary according to the molecule considered (for example, it can depend on the length of the molecule).

Let $\mu_n(x)$ be the expected number of reactions in $\R_{+}(n)$ that molecule $x$ catalyses.  By (R2) we can write this as
$$\mu_n(x) = \PP[(x,r) \in C]\cdot |\R_{+}(n)|,$$ 
for any given $r \in \R_{+}(n)$. 
 
For $\Q(n) = (X(n), \R(n), \gamma_C)$, $F = X(t)$ for some fixed
value of $t$ and  $\Omega \subseteq 2^{X(n)-F}$, let 
$\P_n(\Omega)$  be the probability that $\Q(n)$ has an $\Omega$--complex RAF.
We can now state the first main result of this paper.

\begin{theorem} \label{thm:random_nets}
\mbox{}
Consider a random catalytic reaction system $\Q(n)$ satisfying {\rm (R1)} and {\rm (R2)} and with $F = X(t)$ for a fixed
value of $t$, with $t<n$.  Let $\lambda \geq 0$ and let  $\Omega \subseteq 2^{X(n)-F}$.
\begin{itemize}
\item[{\rm (i)}]
Suppose that $\mu_n(x) \leq \lambda n$ for all $x \in X(n)$. 
Then 
\[
\P_n(\Omega) \leq 1-\exp(-2\lambda x_t^2(1+O(\frac{1}{n})) \mbox{ } (\rightarrow 0 \mbox{ as }  \lambda \rightarrow 0),
\]
where $x_t$ is defined in (\ref{eq:sizeX}). 
\item[{\rm (ii)}]
Suppose that $\mu_n(x) \geq \lambda n$ for all $x \in X(n)$, or that $\mu_n(x)
\geq \lambda \theta_n|x|$ for all $x \in X(n)$, where $\lambda > \log_e(\kappa)$ and where
$\theta_n = \frac{1}{\kappa}(1+\frac{n\kappa^{n+1}}{r_n}) \sim 1$. Then,
\[
\P_n(\Omega) \geq 1 - \frac{\kappa (\kappa e^{-\lambda})^t}{1 - \kappa e^{-\lambda}} \mbox{ } (\rightarrow 1 \mbox{  as }  \lambda \rightarrow \infty).
\]
\end{itemize}
\end{theorem}

To illustrate  Theorem~\ref{thm:random_nets} consider binary sequences, and a food set consisting of the 6 molecules of length at most 2 (thus $\kappa=t=2$, which was the default setting for the simulations in Hordijk and Steel 2004). Then taking $\lambda=4$ in Theorem \ref{thm:random_nets}(ii) we have $\P_n >0.99$.

As an immediate corollary of Theorem~\ref{thm:random_nets} we obtain the following result, which confirms the two 
conjectures posed in Steel (2000).

\begin{corollary}
\label{maincor}
\mbox{}
Consider random catalytic reaction systems $\Q(n)$ ($n \geq t$)
satisfying {\rm (R1)} and {\rm (R2)} and with $F = X(t)$ for a fixed
value of $t$. Take $\Omega = \emptyset$, and let $\P_n =
\P_n(\emptyset)$, the probability that $\Q(n)$ has a RAF. 
\begin{itemize}
\item[{\rm (i)}]
If $$\max_{x \in X(n)}\frac{\mu_n(x)}{n}  \rightarrow 0 \mbox{ as $n \rightarrow \infty$}$$
then $\lim_{n \rightarrow \infty} \P_n = 0$. 
\item[{\rm (ii)}]
If $$\min_{x \in X(n)}\frac{\mu_n(x)}{|x|}  \rightarrow \infty \mbox{ as $n \rightarrow \infty$}$$
then $\lim_{n \rightarrow \infty} \P_n = 1$. 
\end{itemize}
\end{corollary}

{\bf Remarks}
\begin{itemize}
\item Corollary~\ref{maincor} has been worded in such a way that it clearly remains true if we interchange the terms
$\frac{\mu_n(x)}{|x|}$ and $\frac{\mu_n(x)}{n}$ in either part (i) or part (ii) or both. 
\item
The condition described in Corollary~\ref{maincor}(ii) suffices to guarantee (for large $n$) a RAF involving 
all the molecules in $X(n)$. However it does not guarantee that all of $\R_{+}(n)$ is an $RAF$. The condition for
this latter event to hold with high probability as $n\rightarrow \infty$
(assuming for simplicity that $\mu(x)$ is constant, say $\mu_n$, over
$X(n)$) is the stronger condition that $$\liminf_{n \rightarrow
  \infty}\frac{\mu_n}{n^2} > \log_e(\kappa).$$  This follows from (a slight extension of) Theorem 1
of Steel (2000).
\item
Note that if we were to view a sequence $(x_1, x_2, \ldots x_n) \in
X(n)$ and its reversal $(x_n, x_{n-1}, \ldots, x_1)$ as equivalent
molecules then Corollary~\ref{maincor} still holds since asymptotically
(with $n$) palindromic sequences have a negligible influence in the
calculations.
\item
Similarly, if we were to modify (R2) to 
require that  any molecule $x$ cannot catalyse a reaction $r$ for which $x$
is a reactant, then  Corollary 4.2 would still hold (and Theorem
\ref{thm:random_nets} would only be slightly modified) since the number of
reactants in any reaction $r$  is asymptotically negligible (with $n$) compared
with the total number of molecules that could catalyse $r$.
\item
Note that the lower bound on $\P_n$ in Theorem~\ref{thm:random_nets}(ii) is valid for any value of $n > t$ (previous studies, from Kauffman's (1986) onwards, had drawn conclusions by considering limits as $n$ tended to infinity, but the bound in Theorem~\ref{thm:random_nets}(ii) is independent of $n$). Thus, very large systems are not necessarily required for self-sustaining random autocatalysis, a concern that had been raised by Szathm{\'a}ry (2000). 
\end{itemize}

To establish Theorem \ref{thm:random_nets} we require first two further results - Lemma 
\ref{lem:global_cat} and Proposition~\ref{prop:gen_random_nets}, and to describe them we 
introduce a further definition. 

We say that a reaction $r \in \R(n)$ is {\em globally-catalyzed} (or GC) if there exists any 
molecule in $X(n)$ that catalyzes $r$. By the assumptions (R1) and (R2) above the 
probability that any forward reaction $r$ is GC does not depend on $r$. Let $p_{\ast}$ denote 
this probability and let $q_{\ast} = 1- p_{\ast}$.

We will show that when $p_{\ast}$ is sufficiently large then 
there exists a RAF $\R \subseteq \R_{+}(n)$ such that $X(n)-F \subseteq
\pi(\R)$ - 
in other words all molecules that are not already supplied by $F$ can be generated by catalyzed 
reactions. 

On the other hand, we will show that when $p_{\ast}$ is small enough then 
the probability that there exists any globally catalyzed 
reaction that generates any molecule from $X(t+1)$ from molecules in $X(t)$ 
is small - thus proving that the probability that a RAF exists is small. 

The first step is to estimate the probability of global catalysation.

\begin{lemma} \label{lem:global_cat}
\mbox{}
Consider the system $\Q(n)$ satisfying properties {\rm (R1)} and {\rm (R2)} and 
with $F=X(t)$ for a fixed $t$, and let $\lambda>0$ be any positive constant.
\itemize
\item[{\rm (i)}]
The probability $q_{\ast}$ that a reaction $r \in \R_{+}(n)$ is not globally catalyzed is given by
\[
q_{\ast} =  \prod_{x \in X(n)} \left( 1 - \frac{\mu_n(x)}{r_n} \right).
\]
In particular, 
\item[{\rm (ii)}]
if $\mu_n(x) \leq \lambda n$ for all $x$ then 
\[
q_{\ast} \geq  \exp(-\lambda(1 + O(\frac{1}{n})));
\]
\item[{\rm (iii)}]
if $\mu_n(x) \geq \lambda n$ for all $x$ then 
\[
q_{\ast}  <  e^{-\lambda}.
\]

\item[{\rm (iv)}]
if $\mu_n(x) \geq \lambda\theta_n|x|$ for all $x$ (where $\theta_n$ is as defined in Theorem \ref{thm:random_nets}) then 
\[
q_{\ast}  <  e^{-\lambda}.
\]

\end{lemma}
\begin{proof}
Part (i) is immediate from (R1) and (R2). Part (ii) follows by combining part (i) and 
(\ref{eq:size_ration1}) to give:
$$q_{\ast} \geq \left(1-\frac{\lambda n}{r_n}\right)^{x_n}  \geq (1-\frac{\lambda}{x_n(1-O(\frac{1}{n}))})^{x_n} 
= \exp(-\lambda(1+ O(\frac{1}{n}))).$$  
Part (iii) follow from part (i) together with (\ref{eq:size_ration1}) which gives
$$q_{\ast} \leq \left(1 - \frac{\lambda n}{r_n} \right)^{x_n} \leq \left(1 - \frac{\lambda}{x_n} \right)^{x_n} < e^{-\lambda},$$
as required.
For part (iv), combine part (i), the identity $|\{x \in X(n):
|x|=s\}|=\kappa^s$, and the inequality $(1-a)^b \leq \exp(-ab)$ for $a,b>0$,
to obtain
$$q_{\ast} \leq \prod_{s=1}^n\left(1 - \frac{\lambda s}{r_n} \right)^{\kappa^s} 
\leq \prod_{s=1}^n \exp(-\frac{\lambda s\kappa^s}{r_n}) = \exp(-\frac{c}{r_n}\sum_{s=1}^n s\kappa^s).$$ Now,  $\sum_{s=1}^ns\kappa^s = r_{n+1}/\kappa$ from (\ref{eq:sizeR}), and part (iv) now follows by identifying $\theta_n$ with $\frac{r_{n+1}}{r_n\kappa}$ (again using (\ref{eq:sizeR})). Note that 
$\theta_n$ converges to $1$ as $n \rightarrow \infty$. 
\end{proof}

\begin{proposition} \label{prop:gen_random_nets}
Consider a random catalytic reaction system $\Q(n)$ satisfying properties {\rm (R1)} and {\rm (R2)} and 
with $F=X(t)$ for a fixed $t$, where $t<n$.
As before, denote the probability that a forward reaction is not globally catalyzed by $q_{\ast}$.  Then 
\begin{itemize}
\item[{\rm (i)}]
The probability that $\Q(n)$ has a RAF is at most
$1 - q_{\ast}^{2 x_t^2}$
\item[{\rm (ii)}]
If $\kappa q_{\ast} < 1$ then the 
probability that $\Q(n)$ has a RAF $\R$ with $X(n)-F \subseteq \pi(\R)$ is at least 
\[
1 - \frac{\kappa (\kappa {q_\ast})^t}{1 - \kappa q_{\ast}}.
\]
\end{itemize}
\end{proposition}

\begin{proof}
{\em Part (i).}
Note that there are at most $2 x_t^2$ forward reactions whose reactants (inputs) lie 
in $X(t)$. 
With probability $q_{\ast}^{2 x_t^2}$ none of these reactions is GC, 
in which case there is no RAF for the system. The first part of the proposition now 
follows. 

{\em Part (ii).}  
Note that, for any $s \geq t$  the probability that a molecule $x$ with $|x| = s+1$ is 
not generated by any forward GC reaction from $X(s)$ is given by 
$q_{\ast}^s$. Therefore the expected number of molecules $x$ with
$|x| = s+1$ which are not generated by a forward GC reaction is 
$\kappa^{s+1} q_{\ast}^s$. In particular the probability that there is a molecule in $X(s+1)$ that is not generated
by a forward reaction from $X(s)$ is at most $\kappa^{s+1} q_{\ast}^s$. This in turn implies that the 
probability that all molecules in $X(n)$ are generated by forward GC reactions
is at least 
\[
1 - \kappa \sum_{s=t}^{n} (\kappa q_{\ast})^s \geq  
1 - \kappa \sum_{s=t}^{\infty} (\kappa q_{\ast})^s = 
1 - \frac{\kappa (\kappa {q_\ast})^t}{1 - \kappa q_{\ast}}.
\]
Finally, note that if all molecules in $X(n)-F$ are generated by a set $\R$ of forward GC reactions, and since $t<n$ (so that condition (iii) in the definition of an RAF is satisfied) we have that $\R$ is a RAF for $\Q(n)$.
\end{proof}

{\em Proof of Theorem~\ref{thm:random_nets}}

{\em Part (i)}.
By Proposition~\ref{prop:gen_random_nets} (i) the probability that
$\Q(n)$ has a RAF is at most
$1-q_{\ast}^{2x_t^2}$ which by  Lemma \ref{lem:global_cat}(ii) is at most 
$$1-[\exp(-\lambda(1+O(\frac{1}{n})))]^{2x_t^2} =
1-\exp(-2\lambda x_t^2(1+O(\frac{1}{n}))).$$ 
Clearly if $\Q(n)$ has no RAF, then it also has no $\Omega$--complex
RAF, for any $\Omega~\subseteq~2^{X(n)-F}.$

{\em Part (ii)} This follows, by combining Proposition~\ref{prop:gen_random_nets} (ii) with
Lemma \ref{lem:global_cat} parts (iii) and (iv), and noting that a RAF
$\R$ of $\Q(n)$ for which $X(n)-F \subseteq \pi(\R)$ is also an
$\Omega$--complex RAF for any $\Omega \subseteq 2^{X(n)-F}.$
\qed

\section{An analogous result for CAFs}
\label{seccaf}

The degree of catalysation required for a CAF to arise in the system $\Q(n)$ is much greater
than for a RAF. This seems reasonable since the definition of a CAF involves a 
much stronger requirement than a RAF on a set of reactions. However the extent of the
difference is interesting, and is given by the following analogue of Theorem
\ref{thm:random_nets}. 

\begin{theorem}\label{thm:randomnets2}
\mbox{}
Consider the random catalytic reaction system $\Q(n)$ and suppose that $F = X(t).$ Let $\lambda\geq 0$ and let
 $\Omega \subseteq 2^{X(n)-F}$.

\begin{itemize}
\item[{\rm (i)}]
If
\[
\mu_n(x) \leq \frac{\lambda}{x_t^3}\cdot r_n
\]
for all $x \in X(n)$, then the probability that $\Q(n)$ has a $\Omega$-complex  CAF is at most 
$$1-(1-\frac{\lambda}{x_t^3})^{2x_t^3} \leq 2\lambda.$$ 
\item[{\rm (ii)}]
If
\[
\mu_n(x) \geq \frac{\lambda}{x_t}\cdot r_n,
\]
 for all $x \in X(n)$, then the probability that $\Q(n)$ has a $\Omega$-complex CAF is at least
\[
1 - \frac{\kappa (\kappa e^{-\lambda})^t}{1 - \kappa e^{-\lambda}}.
\]
\end{itemize}
\end{theorem}

Before presenting the proof of this result, we note that while the degree of catalysation required for the likely occurrence of a RAF was
that $\mu_n(x)$ should  grow at least linearly with $n$ (Theorem~\ref{thm:random_nets}) the requirements for a CAF
are quite different: by Theorem~\ref{thm:randomnets2} 
$\mu_n(x)$ must grow at least linearly with $r_n$ - and thereby {\em exponentially} with $n$.

{\em Proof of Theorem~\ref{thm:randomnets2}}
{\em Part (i).}
Let $\R':=\{r \in \R_{+}(n): \rho(r) \subseteq F\}$, the set of all forward reactions that have all their reactants in $F$.
The probability that any given reaction $r \in \R'$ is not catalyzed by at least one element of $F$ is 
given by 
\[
\prod_{x \in F}(1 - \frac{\mu_n(x)}{r_n}).
\]

Thus the probability that none of the reactions in $\R'$ are catalyzed by at least one
element of $F$ is 
\[
(\prod_{x \in  F}(1 - \frac{\mu_n(x)}{r_n}))^{|\R'|}.
\]
In particular if $\mu_n(x) \leq \frac{\lambda r_n}{x_t^3}$,
then, since $|F| = x_t$ and $|\R'| = 2x_t^2$, this probability (that none of the reactions in $\R'$ is
catalyzed by at least one element of $F$) is at least
\begin{equation}
\label{lowerboundseq}
(1 - \frac{\lambda}{x_t^3})^{2x_t^3} \geq 1-2\lambda.
\end{equation}
However when none of the reactions in $\R'$ is catalyzed, then $\Q(n)$
does not have a CAF. Thus the probability that $\Q(n)$ has a CAF is at
most 1 minus the expression in (\ref{lowerboundseq}), as required. 

{\em Part (ii).}  For every molecule in $x \in X(n)$, and each $s \in
\{t, \ldots, n\}$ let $E_s(x)$ be the event that there is at least one
reaction $r_x$ of the form $a+b \rightarrow x$, where $a,b \in X(s)$, 
that is catalysed by at least one molecule in $X(s)$.

Now, if $\mu_n(x) \geq \frac{\lambda r_n}{x_t}$, then for
 any forward reaction $r$, the probability that $r$ is not
catalysed by at least one molecule in $X(s)$ (for $s \geq t$) is at most
$$(1-\frac{\lambda}{x_t})^{x_s} \leq \exp(-\lambda\frac{x_s}{x_t}) \leq e^{-\lambda}$$
and since, for each $x$ 
 there are $|x|-1$ choices for $r_x$ we have
\begin{equation}
\label{bonfereq}
\PP(E_s(x)^c) \leq  \exp(-\lambda(|x|-1)),
\end{equation}
where $E_s(x)^c$ is the complementary event to $E_s(x)$. 

Consider the event $$E_s: = \bigcap_{x \in X(s+1)\setminus X(s)}E_s(x).$$
By (\ref{bonfereq}) and the identity $|X(s+1)-X(s)|=\kappa^{s+1}$ 
we have $P(E_s^c) < \kappa^{s+1}e^{-\lambda s}$
and so
$$\PP(\bigcap_{s=t}^{n-t}E_s) \geq 1- \sum_{s=t}^{\infty} \kappa^{s+1}e^{-\lambda s}
=
1-\frac{\kappa(\kappa e^{-\lambda})^t}{1-\kappa e^{-\lambda}}.$$
However the event $\bigcap_{s=t}^{n-t}E_s$ ensures that 
the nested collection of  reactions $\R_i := \{r_x: x \in X(t+i)\}$, $i =
1, \ldots, n-t$ forms
a CAF for $\Q(n)$, and moreover one for which the maximal set $\R_{n-t}$
generates all elements of $X(n)-F$ - thus it is also a $\Omega$--complex CAF for any
$\Omega \subseteq 2^{X(n)-F}$. This completes the proof. 

\qed

\section{Discussion}
\label{seccon}

The question of how life first arose on earth is a multifaceted problem
that stands out as one of the major questions in science (see for
example Dyson, 1985; Fenchel, 2002; Joyce, 1989; Szmathm{\'a}ry 1999; Szathm{\'a}ry and
Smith, 1997). One dilemma, frequently dubbed the `chicken and egg'
problem is the question of which (if either) 
 came first: hereditary (molecules such as DNA or RNA
that carry information but do not easily catalyse reactions), or
metabolism (proteins that carry out reactions but do not replicate).
An alternative possibility is than an
autocatalytic system of molecules including RNA and proteins and
possibly other molecules emerged as the first primitive prebiotic
system. The theoretical
investigation of catalytic reaction systems is an attempt to address
just one aspect of this theory.   This concerns the issue of whether,
as Kauffman has maintained, we should expect self-sustaining,
autocatalytic networks to emerge in random chemical systems once some
threshold (in `complexity', `connectivity' or `catalysation rate') is
exceeded, or whether there is the requirement of some fine-tuning of the
underlying biochemistry for such networks to occur.  Orgel (1992)
raises this as concern about   autocatalytic network models
 commenting that ``it
is always difficult in such theoretical models to see how to close the
cycle without making unreasonable assumptions about the specificity of
catalysis.'' 

Our results here have helped delineate precisely how much catalysation
is required in order for random sequence-based chemical reaction systems (without any
`fine-tuning') to likely give rise to a RAF. In contrast to a CAF, where
a high degree of catalysation is required when the maximal sequence
length $n$ is large,  the likely occurrence of a RAF depends just on
whether the catalysation function $\mu_n(x)$ grows sublinearly or
superlinearly with $n$ (Corollary~\ref{maincor}). The techniques
developed in this paper may provide some analytical predictive tools 
for biochemists design {\em in vitro} prebiotic experiments
with  large number of variants of RNA sequences and other molecules.

The development of  a self-sustaining autocatalytic system would clearly
be only one step towards life, in particular a reproducing 
system that is capable of
undergoing Darwinian selection eventually needs to develop. Here the recent
concept of a `Eigen-Darwin' cycle (Poole et al. 1999) may hold promise.

 Questions for future work would be to explore how the results in
 this paper would be influenced by allowing random inhibitory
 catalysations, or side reactions that could destroy some of
 the crucial reactants (this problem has been referred to by
 Szathm{\'a}ry (2000) as the ``plague of side reactions'').  
This second phenomena can be formally regarded as a special case of the first,
 since if $x$ is a reactant for 
 a reaction $r$ and $x$ is degraded in the presence of
another molecule $y$ then we can (formally) 
regard $y$ as inhibiting the reaction $r$.  The model studied in this paper
 could also be refined to better suit the graph-theoretic properties of real metabolic networks which have recently been investigated (Jeong {\em et al.}, 2000; Wagner and Fell, 2001).

\section{Acknowledgements}
The first author is supported by a Miller fellowship. 
The second author thanks the NZIMA (Maclauren Fellowship) for supporting this work. We also thank the referees for helpful comments on an earlier version of this manuscript.

{}

\section{Appendix: Proof of Proposition~\ref{hard}}

The decision problem is clearly in the class NP.  To show it is NP-complete we 
provide a reduction from $3$--SAT.
Consider an expression  $P$ in conjunctive normal form involving binary variables $x_1, \ldots x_n$ and
where each clause in $P$ involves at most three variables. 
Thus we can write
\[
P= C_1 \wedge C_2 \wedge \cdots \wedge C_k
\] 
where 
\[
C_i= \vee_{j \in T(i)}x_j \vee_{j \in F(i)} \overline{x}_j
\]
and $T(i), F(i) \subseteq \{1, \ldots, n\}$, $|T(i)|+|F(i)|=3$.

Given $P$ construct a catalytic reaction system $\Q=(X,\R, \gamma_{C(+),C(-)})$ as follows: let $F:=\{x_1, \ldots, x_n\}$, let 
$$X := 
\{x_1,\ldots, x_n,f_1,\ldots,f_n,t_1,\ldots,t_n,\theta_1,\ldots,\theta_k,1\}.
$$ 
Informally, $x_i$ will correspond to the variable $x_i$ in the formula; a reaction producing
$t_i$ (respectively $f_i$) will be catalyzed if the truth assignment of $x_i$ is true (respectively false), and the reaction producing $\theta_i$ will be catalyzed if the $i$'th clause 
is satisfied. 

More formally we let $\R = \R_1 \cup \R_2 \cup \R_3$ where $\R_1$ consists of 
all reactions 
\[
\begin{array}{ll}
x_i \xrightarrow{1(+), t_i(-)} f_i, & 
x_i \xrightarrow{1(+), f_i(-)} t_i, 
\end{array}
\]
for $1 \leq i \leq n$. In words $x_i$ is the sole reactant for $f_i$ and $t_i$ but 
$f_i$ inhibits the catalysation of $t_i$ and vice-versa.

$\R_2$ consists of all reactions 
\[
\left\{
\begin{array}{ll} 
t_j \xrightarrow{1(+)} \theta_i & \mbox{ if } j \in T(i),\\ 
f_j \xrightarrow{1(+)} \theta_i & \mbox{ if } j \in F(i),\\ 
\end{array}\right.
\]
for $1 \leq j \leq n$ and $1 \leq i \leq k$.
Finally $\R_3$ consists of the single reaction
\[
\{\theta_1,\ldots,\theta_k\} \xrightarrow{1(+)} 1.
\]

Now we claim that $P$ has a satisfying truth assignment if and only if $\Q$ has a RAF. To establish this, first
assume that $P$ has a satisfying assignment. Fix such an assignment $z$ and 
let $\{T,F\}$ be  a partition of $\{1,\ldots, n\}$ corresponding 
to the variables that are true (respectively false) in $z$.

Now consider $\R'_1 \cup \R'_2 \cup \R_3$ where 
$\R'_1 \subseteq \R_1$ consists of the  reactions $x_i \rightarrow t_i$ for all $i \in T$, and the reactions 
$x_i \rightarrow f_i$ for all $i \in F$. $\R'_2$ will consist of the reactions
$t_i \to \theta_j$ for all $i \in T \cap T(j)$ and $f_i \to \theta_j$ for all 
$i \in F \cap F(j)$. Since the assignment $z$ satisfies the formula it 
follows that $\R'_1 \cup \R'_2 \cup \R_3$ is a RAF. 

Next we have to show that if the system has a RAF the formula has a 
satisfying truth assignment. Suppose the system has a RAF $\R'$. Clearly 
$\R_3 \subset \R'$. This in turn implies that the reactions producing $\theta_1,\ldots,\theta_k$ are 
all catalyzed. Thus for all $1 \leq i \leq k$, there either exists some 
$j \in T(i)$ such that the reaction producing $t_j$ is catalyzed or there exists some $j \in F(i)$ such 
that the reaction producing $f_j$ is catalyzed. Moreover, for all $i$ at most one of the reactions producing 
$t_i$ and
$f_i$ can be catalyzed. We now define $z_i$ to be true if the reaction producing $t_i$ is catalyzed 
and false if the reaction producing $f_i$ is catalyzed ($z_i$ is defined arbitrarily otherwise). 
Then $z$ is a satisfying assignment as required.

\end{document}